\documentstyle[11pt,newpasp,twoside]{article}
\def\edcomment#1{\iffalse\marginpar{\raggedright\sl#1\/}\else\relax\fi}
\reversemarginpar
\def\ltsima{$\; \buildrel < \over \sim \;$}
\def\lesssim{\lower.5ex\hbox{\ltsima}}
\def\gtsima{$\; \buildrel > \over \sim \;$}
\def\gtrsim{\lower.5ex\hbox{\gtsima}}

\begin{document}
\title{GRBs in Pulsar Wind Bubbles: Observational implications}
\author{Dafne Guetta}
\affil{Osservatorio Astrofisico di Arcetri, L.go E. Fermi 5, 501225 Firenze}
\author{Jonathan Granot}
\affil{Institute for Advanced Study, Olden Lane, Princeton, NJ 0854}

\begin{abstract}
We present the main observational features expected for Gamma-Ray
Bursts (GRBs) that occur inside pulsar wind bubbles (PWBs). This is
the most natural outcome of the supranova model, where initially a
supernova (SN) explosion takes place, leaving behind a supra-massive
neutron star, which loses its rotational energy over a time
$t_{\rm sd}$ and collapses to a black hole, triggering a GRB.
We find that the time delay, $t_{\rm sd}$, between the
SN and GRB events is the most important parameter that
determines the behavior of the system. We consider the 
afterglow, prompt GRB and PWB emission. Constraints on the
model are derived for a spherical PWB, from current afterglow 
observations and the lack of direct detection of the PWB emission.
We find that a simple spherical model
cannot account for the X-ray features detected in some afterglows, 
together with the typical afterglow emission that was observed for the 
same GRBs. The discrepancies with observations may be 
reconciled by resorting to a non-spherical geometry,
where the PWB is elongated along the polar axis. 
Finally, we predict that the inverse
Compton upscattering of PWB photons by the relativistic
electrons of the afterglow (external Compton) should lead to
high energy emission during the early afterglow that may explain
the GeV photons detected by EGRET for a few GRBs, and should be
detectable by future missions such as GLAST.
\end{abstract}


\section{Introduction}

Despite the large progress in GRB research over 
the last few years, the identity of their progenitors is still 
perhaps the most interesting open question.
GRB Progenitor models are divided into two main categories:
(i) a merger of two compact stars, 
(ii) the death of a massive star. 
Short GRBs (lasting $\lesssim 2\;$s) are usually 
attributed to the category (i),
while long GRBs are attributed to category (ii),
which includes the failed SN (Woosley 1993)
or hypernova (Pacy\'nski 1998) models, and the supranova model 
(Vietri \& Stella 1998), where a massive
star explodes in a SN leaving behind a supra-massive neutron 
star (SMNS) which after a time delay, $t_{\rm sd}$, loses its rotational 
energy and collapses to a black hole (BH),
triggering the GRB. All these models have the same final stage
(large accretion rate onto a newly formed BH) and a similar energy budget 
($\lesssim 10^{54}\;{\rm ergs}$).

This work concentrates on the supranova model,
focusing on its possible observational signatures. 
The original motivation for this model was to provide a relatively baryon clean
environment for the GRB jet. As it turned out, it also seemed to
naturally accommodate the later detection of iron lines in several
X-ray afterglows (Lazzati, Campana, \& Ghisellini 1999; Piro et
al. 2000; Vietri et al. 2001). Recently, it has been suggested that
the most natural mechanism by which the SMNS can lose its
rotational energy is through a strong pulsar type wind, between
the SN and the GRB events, which creates a pulsar
wind bubble (PWB), also referred to as a plerion (K\"onigl \&
Granot 2002, KG hereafter; Inoue, Guetta \& Pacini 2002).
Here we report the main results of Guetta \& Granot (2002a, GG hereafter),
and refer the reader to this work for more details. 

The  most important parameter that determines the behavior of the 
system is the time delay, $t_{\rm sd}$, between the SN and GRB events. 
The value of $t_{\rm sd}$ is set by the timescale on which 
the SMNS loses its rotational energy due to magnetic dipole 
radiation (see Eq. 2 of GG) and depends mainly on the strength of 
the SMNS magnetic field; it can ranges anywhere between weeks to years.
Another important parameter is the Lorentz factor, $\gamma_w$, of
the pulsar wind, emanating from the SMNS, which is
expected to be in the range $\sim 10^4-10^7$ (KG).
An important difference between our analysis and previous works
(KG; Inoue, Guetta \& Pacini 2002) is that we allow for a proton
component in the pulsar wind, that carries a significant fraction
of its energy. In the standard model, the external medium is
composed of cold protons and electrons (in equal numbers), and has
a density profile that scales with the distance from the source as
$r^{-k}$, where $k=0$ for an ISM and $k=2$ for a stellar wind. In
our scenario, the external medium is made up of hot protons and
cold $e^\pm$ pairs, where there are $\sim 10^3$ times more pairs
than protons. Nevertheless, the protons hold most of the energy in 
the PWB due to their large internal energy, which also dominates the
effective density that  is responsible for the deceleration of the
afterglow shock. The value of $k$ for our model ranges between
$k=0$, that is similar to an ISM, and $k=1$, that is intermediate
between an ISM and a stellar wind (KG).

\section{The Behavior of the System for Different Time Delays}
\label{t_sd}

In this section we go over the main observational signatures of the PWB model,
following the different regimes in $t_{\rm sd}$ (for the
detailed calculations on how these results are derived we refer the 
reader to GG):

\noindent 1. For extremely small values of 
$t_{\rm sd}<t_{\rm col}=R_\star/\beta_b c\approx
0.9\,R_{\star,13}\beta_{b,-1}^{-1}\;{\rm hr}$, 
where $R_\star=10^{13}R_{\star,13}\;{\rm cm}$
is the radius of the progenitor star (before it explodes in a SN), 
the stellar envelope does not have enough time to increase its radius 
considerably before the GRB  goes off, and the supranova model reduces 
to the collapsar model. In this respect, the 
collapsar model may be seen as a special case of the supranova model. 

\noindent 2. If $t_{\rm sd}<t_\tau\sim 0.4\;{\rm yr}$, the SNR shell is 
optically thick to Thomson scattering, and the radiation from the
plerion, the prompt GRB and the afterglow cannot escape and reach
the observer. If the SNR shell is clumpy (possibly due to the
Rayleigh-Taylor instability, see \S 2 of GG), then the Thomson
optical  depth in the under-dense regions within the SNR shell may
decrease below unity at $t_{\rm sd}$ somewhat smaller than
$t_\tau$, enabling some of the radiation 
to escape. The only signatures that we expect for this range of
$t_{\rm sd}$ are the neutrino emission due to p-p collisions or
photo-meson interactions, and high energy photons above
$0.5\,(t_{\rm sd}/t_\tau)^{-2}\;$MeV  
(Guetta \& Granot 2002b and Granot \& Guetta 2002b).

\noindent 3. For $t_\tau<t_{\rm sd}<t_{\rm Fe}\sim 1\;{\rm yr}$ 
the SNR shell has a Thomson optical depth smaller than unity, 
but the optical depth for the iron line features is still $\gtrsim 1$ 
so that detectable X-ray line features, like the iron 
lines observed in several afterglows, can be produced. In this range of 
$t_{\rm sd}$ we expect a very large effective density 
($\sim 10^5\;{\rm cm^{-3}}$) and electron 
number density ($\sim 10^3\;{\rm cm^{-3}}$). 
This effects the afterglow emission in 
a number of different ways: i) The self absorption frequency of the 
afterglow is typically above the radio, implying no detectable radio 
afterglow, while radio afterglows were detected for GRBs 970508, 970828, 
and 991216, where the iron line feature for the latest of these three is 
the most significant detection to date 
($\sim 4\sigma$, Piro et al. 2000). We also typically 
expect the self absorption frequency of the plerion emission to be above 
the radio in this case, so that the radio emission from the plerion 
should not be detectable, and possibly confused with that of the afterglow. 
ii) A short jet break time $t_j$ and 
a relatively short non-relativistic transition time $t_{\rm NR}$ are 
implied, as both scale linearly with $t_{\rm sd}$ and are in the right 
range inferred from observations 

\noindent 4. Finally, for $t_{\rm sd}>t_{\rm Fe}$, we expect no iron lines.
When $t_{\rm sd}$ is between $\sim 2\;{\rm yr}$ and $\sim 20\;{\rm
yr}$ the radio emission of the plerion may be detectable for
$\gamma_w\lesssim 10^5$. The lack of detection of such a radio
emission excludes values of  $t_{\rm sd}$ in this range, if indeed
$\gamma_w\lesssim 10^5$, as is needed to obtain reasonable values
for the break frequencies of the afterglow (see Fig.1 of GG and the relevant
discussion there). 

\section{High energy emission}

An interesting new ingredient of the PWB model, is that the GRB
and its afterglow occur inside a photon rich plerionic environment.
These photons can be upscattered by the relativistic electrons behind
the afterglow shock, producing a high energy emission (external Compton, EC). 
Fig. 2 of GG and Fig. 1 of Granot \& Guetta (2002a) 
show that the EC emission can account for the
high energy emission detected by EGRET for GRB 940217 (Hurley et
al. 1994), and is consistent with the flux level and 
moderate time decay observed in this case.

\section{Conclusions}

In this work we have presented the main results obtained in GG
and Granot \& Guetta (2002a) regarding the  observational implications 
for GRBs that occur inside pulsar wind bubbles (PWBs), as expected in 
the supranova model. 
We find that a simple spherical model cannot produce the iron line
features observed in several afterglows together with the other,
more conventional, features of the afterglow emission from these
bursts. However, if the iron lines are not real, then a simple
spherical model can explain all other observations for $t_{\rm
sd}\gtrsim 20\;$yr. The latter is required in order to explain
typical afterglow observations and the lack of direct detection 
of the plerion emission in the radio during the afterglow.

If the iron line detections are real, then in the context of the
PWB model, this requires deviations from a simple spherical
geometry. The most straightforward variation of the simple model
is a PWB that is elongated along its polar axis. Such a geometry
may arise naturally within the context of this model (KG; GG).
With an elongated geometry, the PWB model can account for all the
observed features in the afterglow, and it offers a number of
advantages in comparison to other models: 
i) It provides a relatively baryon clean environment
for the GRB jet, which is required in order to produce a highly
relativistic outflow. This arises as the initial SN expels
most of the stellar envelope to a large distance from the site of
the GRB, and the strong pulsar wind effectively sweeps up the remaining 
baryonic matter. ii) An important advantage of this
model is that it can naturally explain the large values of
$\epsilon_B$ and $\epsilon_e$ that are inferred from fits to
afterglow data (KG), thanks to the large magnetic fields in the
PWB and the large relative number of electron-positron pairs. This
is in contrast with standard environment that is usually assumed
to be either an ISM or the stellar wind of a massive progenitor,
that consists of protons and electrons in equal numbers. In this 
case, the pre-existing magnetic field, that is amplified due to the 
compression of the fluid in the shock, is too small to explain the values 
inferred from afterglow observations, and further magnetic 
field amplification or generation at the shock is required. 
iii) All the detections of GRB afterglows to date 
are for the long duration sub-class of GRBs (with a duration $\gtrsim
2\;{\rm s}$), that are believed to arise from a massive star
progenitor, which according to the collapsar model should imply a
stellar wind environment ($k=2$). However, a homogeneous external
medium ($k=0$) provides a better fit to the observational data for
most GRB afterglows. This apparent contradiction is naturally
explained in the context of the PWB model, where $k$ ranges
between $0$ and $1$, while we still have a massive star
progenitor. 

Another advantage of the PWB model is its
capability of explaining the high energy emission observed in some
GRBs. We find that the high energy emission
during the early afterglow at photon energies $\gtrsim 100\;$keV
is dominated by the EC component. We predict that such a high
energy emission may be detected in a large fraction of GRBs with
the upcoming mission GLAST. 


\end{document}